# Symmetry Strategy for Rapid Discovery of Abundant Fractional Quantum Ferroelectrics

Guoliang Yu[1,2,†], Junyi Ji[1,2,†], Changsong Xu[1,2]*, and H. J. Xiang[1,2,3]*

[1]*Key Laboratory of Computational Physical Sciences (Ministry of Education), Institute of Computational Physical Sciences, State Key Laboratory of Surface Physics, and Department of Physics, Fudan University, Shanghai 200433, China*

[2]*Shanghai Qi Zhi Institute, Shanghai 200030, China*

[3]*Collaborative Innovation Center of Advanced Microstructures, Nanjing 210093, China*

[†] *These authors contributed equally to this work.*
Email: csxu@fudan.edu.cn, hxiang@fudan.edu.cn

**Abstract**

Traditional ferroelectrics are limited by Neumann's principle, which confines exploration of ferroelectrics within polar point groups. Our recent work [Nat. Commun. 15, 135, (2024)] proposes the concept of fractional quantum ferroelectricity (FQFE) that extend the playground of ferroelectricity to non-polar point groups. Here, we apply group theory and introduce an efficient symmetry strategy to identify FQFE candidates. Integrated with a high-throughput screening scheme, we go through 171,527 materials and identify 202 potential FQFE candidates, which are already experimentally synthesized. In addition, we point out that the essence of FQFE is fractional atomic displacements with respect to lattice vectors, which can actually result in both fractional (type-I) and integer (type-II) quantized polarization, respectively. Through performing first-principles calculations, we verify the symmetry-predicted switchable FQFE properties in bulk $AlAgS_2$ and monolayer $HgI_2$. Notably, $AlAgS_2$ exhibits an ultra-low switching barrier of 23 meV/f.u. and interlocked in-plane/out-of-plane polarization, while $HgI_2$ demonstrates large spontaneous polarization of 42 $\mu C/cm^2$. Our findings not only advance the understanding on FQFE, but also offer guidance for experimental exploration and design of novel ferroelectric materials.

**Introduction**

Ferroelectric (FE) materials, with their electrically switchable polarization states, are essential for various applications in energy-efficient nanoelectronic devices and storage [1-5]. For conventional FEs, the polarization arises from small atomic displacements, which corresponds to the variants in Wyckoff positions, such as the $z$ component of $(0, 0, z)$. In such cases, the direction of polarization must be consistent with the symmetry of the FE phase, as required by Neumann's principle [6], which limits the exploration of ferroelectricity to the 10 polar point groups out of the total 32 ones [7-10].

Recently, we proposed a new type of ferroelectricity, the fractional quantum ferroelectricity (FQFE) [11], which display unconventional polarization arising from fractional atomic displacements with respect to lattice vectors (see Fig. 1). Such quantum refers to the polarization induced by a unit charge passing through the distance of a lattice vector **a**, as defined $\mathbf{Q} = \frac{e}{\Omega}\mathbf{a}$ ($\Omega$ denotes the volume of the cell) in modern theory of polarization [12,13]. In contrast to conventional FEs, ion(s) in fractional quantum ferroelectrics (FQFEs) displace a large distance among fixed components of Wyckoff positions. For example, monolayer $\alpha$-$In_2Se_3$ crystallizes in $C_{3v}$ point group and was believed to only allow out-of-plane polarization, while the in-plane position of the Se ion moves from (1/3, 2/3) to (2/3, 1/3), which exactly leads to FQFE that explains the observed unexpected in-plane polarization [14-17]. Due to the large atomic displacements, FQFEs intrinsically exhibit large polarizations. According to FQFE theory, novel polarization that contradicts with Neumann's principle can exist in 28 point groups, including 8 polar one and also 20 non-polar ones [11]. Apparently, FQFE largely extend the playground of ferroelectricity and implies novel physics and applications.

Despite the wide theoretical existence of FQFEs, currently known FQFE systems are limited to monolayer $\alpha$-$In_2Se_3$ and bulk zinc-blende AgBr. Moreover, though AgBr is perfect for theoretical demonstration of FQFE in non-polar point group, it is experimentally found unstable in air. The scarcity of FQFE candidates leads to challenges to experimental exploration of the intriguing physics and promising application of FQFE. The scarcity of FQFEs can be attributed to their adherence to symmetry rules that are distinct from common understanding, potentially resulting in the overlooking of possible FQFEs [7-10]. A practical problem is that, if certain structure is assumed to be one state of FQFE, there still lacks of an effective method to determine the other different but symmetrically equivalent state (or other states). It is thus highly desirable to develop efficient method to rapidly identify possible FQFEs.

In this Letter, we develop an efficient symmetry strategy for rapid discovery of FQFEs. Such strategy utilizes group theory analysis and involves lattice symmetry. Applying such strategy, a high-throughput first-principles screening workflow is performed to search potential FQFE candidates. Running through over 171,000 materials, we identify 202 experimentally synthesized materials as FQFE candidates, with 12 candidates in polar point groups and 190 in non-polar ones. Based on the screening results, the FQFEs are classified into type-I and type-II, where the latter actually exhibits integer quantized polarization. Moreover, through further density functional theory (DFT) calculations, bulk $AlAgS_2$ is found to exhibit interlocked in-plane type-I FQFE and out-of-plane conventional ferroelectricity, with an ultra-low switching barrier of 23 meV/f.u.. More interestingly, type-II FQFE is demonstrated in monolayer $HgI_2$ with a non-polar point group and it exhibits a large polarization of 42 $\mu C/cm^2$.

**Classification of the FQFE**

According to the FQFE theory, the atomic displacements should be fractional lattice vectors and the polarization should intuitively be fractional multiples of polarization quantum [11]. Such simple case is termed as type-I FQFE here. As exampled by the tetragonal system in Fig. 1a, it is assumed that the sites (0,1/2,0) and (1/2,1,0) are equivalent. When the $A^+$ ion moves 1/2 of the lattice along the [110] direction from position (0,1/2,0) to (1/2,1,0), it induces a polarization difference $\Delta \mathbf{P} = \mathbf{Q}/2$. However, a different case is that, though displacements are fractional, the polarization is integer times of polarization quantum, which case is termed as type-II FQFE. For instance, as shown in Fig. 1b, $A^{2+}$ ion displaces in the same way with that in Fig. 1a, while the polarization difference yields $\Delta \mathbf{P} = \mathbf{Q}$. Another scenario leading to type-II FQFE is that, though the charge of A is +1, the multiplicity of A ions is an even number. Obviously, type-II FQFEs indicate stronger spontaneous polarization compared to type-I.

**Symmetry strategy and high-throughput screening of FQFE.**

Applying group theory analysis, we develop a symmetry strategy to identify FQFE, as illustrated in Fig. 2a. Specifically, the identification process follows four steps: (i) Given a crystal structure of FE phase $L_1$, find its space group $G$, and the space group $G_L$ of its symmetrized lattice. (ii) Construct all other FE phase structure $L_2$ in the way that $L_2 = hL_1$, where $h$ is a symmetry operation in $G_L/G$ (i.e., belonging to $G_L$ but not to $G$). (iii) For each $L_2$, check whether the phase transition from $L_1$ to $L_2$ is likely to be FQFE. This can be determined by whether $\Delta d_{A,\alpha}$ ($\alpha = x, y, z$), the total displacement of the element A along the $\alpha$ direction, is a fraction M/N. Here, $\Delta d_{A,\alpha} = \Sigma_i d^i_{L2,A,\alpha} - d^i_{L1,A,\alpha} = M/N$ ($\alpha = x, y, z$, $N = 2, 3, 4, 6, 8$; $M = 1, 2, ... N-1$), where $d^i_{L1,A,\alpha}$ and $d^i_{L2,A,\alpha}$ represent the $\alpha$ fractional coordinates of the $i$-th A atom in the $L_1$ and $L_2$ phases respectively. (iv) Output the fraction $\Delta d_{A,\alpha}$ and corresponding $L_2$. For a better understanding of this strategy, let us take CuCl as an example: (i) As shown in Fig. 2b, the space group of $L_1$ phase CuCl is $G = F$-$43m$. Removing all atoms, and the space group of the lattice of the structure is found to be $G_L = Fm$-$3m$ (see Fig. 2b). (ii) $h$ is a symmetry operation belonging to $Fm$-$3m$ but not to $F$-$43m$. Take $h = I$ (inversion) as an example, another FE phase $L_2$ can be constructed as $L_2 = IL_1$. Note that here other choices of $h$ will lead to the same FE phase $L_2$. (iii) As illustrated in Fig. 2b, in the unit cell of $L_1$ and $L_2$, the position of Cu atom at (0,0,0) is invariant under operation $I$, while that of Cl atom is changed under *I*, from (1/4,1/4,1/4) to (3/4,3/4,3/4). As a result, $\Delta d_{Cu,\alpha} = 0$ and $\Delta d_{Cl,\alpha} = (1/2,1/2,1/2)$. Therefore, the CuCl with the *F-43m* space group is a candidate for FQFE (see details in Fig. 1 and Table S1).

To extensively identify FQFEs, besides the presently developed symmetry strategy, the high-throughput screening combined with DFT is also necessary [18-22]. As shown in Fig. 2c, our search starts with over 171,000 experimentally synthesized materials, comprising 154,718 bulk materials from the Materials Project [23] and 16,819 monolayers from the Computational 2D Materials Database (C2DB) [24,25]. Subsequently, we exclude materials with band gaps smaller than 0.1 eV, as these are likely to suffer from leakage. By the end of this step, we have 26176 materials left. Then, we apply the presently developed symmetry strategy, which has been integrated in the PASP software [26]. Finally, the 202 FQFE candidates, comprising 12 and 190 materials with polar and non-polar groups, was further verified through DFT calculations. All obtained bulk and 2D FQFE materials are listed in Table S1 and S2 of the Supplementary Materials (SM) [27], respectively.

Let us take an overlook of the obtain FQFE candidates. In Fig. 2d, we present a summary of the band gap versus polarization for all FQFE candidates. As aforementioned, FQFEs generally exhibit significant spontaneous polarization due to their substantial atomic displacements. For instance, monolayers of $HgI_2$ and $GeS_2$ with the *P-4m2* space group display polarizations of 42.0 and 113.2 μC/cm$^2$, respectively, while the bulk BN, SiC, and CuCl with the *F-43m* space group demonstrate even higher values, reaching up to 631.7, 586.6, and 191.1 μC/cm$^2$, respectively. Among these materials, bulk $BeF_2$ with a space group of *P6$_2$22* exhibits the largest band gap of 8.09 *e*V, accompanied by a spontaneous polarization of 71.4 μC/cm$^2$. We further exam the switching barrier of polarization but did not calculate it for all candidates, as pathway predicted by nudged elastic band (NEB) method may not reflect the real switching process. Instead, we selectively choose ionic systems with simple crystal structure, as they usually correspond to lower switching barriers and the NEB method works well for them (see Table S3). In the following sections, we demonstrate FQFE with the example systems of bulk $AlAgS_2$ and monolayer $HgI_2$, which both exhibit low polarization switch barriers.

**Type-I FQFE in bulk $AlAgS_2$.**

The bulk $AlAgS_2$ has been experimentally synthesized [28] and its structure can be viewed as two $AlS_2$ layers intervened by a layer of Ag ions (see Fig. 3a). The $AlAgS_2$ displays a band gap of 1.71 eV, and the Ag ion exhibit +1 valence state. Firstly, we applied the symmetry strategy to identify the FQFE of $AlAgS_2$. As shown in Fig. 3a, the space group of the $L_1$ phase $AlAgS_2$ is $G = P3m1$, while the space group of its lattice is $G_L = P\text{-}3m1$. Another phase, $L_2$, equivalent to $L_1$, can be constructed by $L_2 = hL_1$, where $h = G_L/G$ represents a symmetry operation. Presently, $h$ can choose from inversion symmetry $I$ centered at any Al atom, and two-fold rotation $C_2$ along the line connecting any Al atom and its nearest Al neighbor, both of which can lead to the same $L_2$ phase. As illustrated in Fig. 3a, under the symmetry operation $h$, the positions of Al and S atoms remain invariant in the $L_1$ and $L_2$ phases, while those of Ag atoms change from (2/3,2/3,$z$) to (1/3,1/3,-$z$). As a result, the displacement of the Al, S, and Ag atoms are $\Delta d_{Al,\alpha} = 0$, $\Delta d_{S,\alpha} = 0$, and $\Delta d_{Ag,\alpha} = (1/3,1/3,2z)$, respectively. Therefore, $AlAgS_2$ emerges as a potential candidate for FQFE owing to the fractional in-plane displacement of $\frac{1}{3}\mathbf{a} + \frac{1}{3}\mathbf{b}$ of the Ag ions.

Further, we validated the in-plane polarization arising from FQFE in bulk $AlAgS_2$ by DFT calculations. Figure 3b shows the evolution of the in-plain polarizations along the $L_1$-TS-$L_2$ pathway (TS stands for transition state) calculated by the Berry phase method [29-31]. One can see that the $L_1$ and $L_2$ phases have in-plane polarizations with amplitudes 21.97 and -21.97 μC/cm², respectively. The polarization quantum $\mathbf{Q}$ of $AlAgS_2$ along the [110] direction is 131.82 μC/cm², which is obtained by $\mathbf{Q} = \frac{e}{\Omega}\mathbf{a}$, where $e$, $\mathbf{a}$, and $\Omega$ represents the electron charge, lattice vector along the [110] direction, and the volume of the unit cell. Therefore, the in-plane polarization of AlAgS$_2$ is type-I FQFE, and the polarization difference between the L$_1$ and the L$_2$ phase is Δ**P** = **Q**/3. Furthermore, to further demonstrate the ferroelectricity of AlAgS$_2$, we show that the polarization can be readily switched by the application of an external electric field. Figure 3a also shows the minimum energy pathway of two low-symmetry phase L$_1$ and L$_2$ using the climbing image nudged elastic band (CI-NEB) method [32]. One can see that the transforming between L$_1$ and L$_2$ phases only needs to overcoming an energy barrier of about 22 meV/f.u., which is much smaller than the 200 meV/f.u. in the experimentally confirmed FE PbTiO$_3$ [33]. This ultra-low polarization switched barrier indicating that the FQFE of AlAgS$_2$ is likely to be experimentally confirmed. Therefore, these results determine that bulk AlAgS$_2$ has switchable in-plane type-I FQFE.

In addition, as shown in Fig. 3a, a difference exists in the distance between the central Ag atomic layer and the two S atomic layers in AlAgS$_2$, leading to inversion symmetry breaking and inducing conventional out-of-plane ferroelectricity that satisfies Neumann's principle. The out-of-plane electric dipole moment of AlAgS$_2$ is 15.55 μC/cm$^2$ (see Fig. 3b). Interestingly, the out-of-plane and in-plane polarization of AlAgS$_2$ are interlocked, akin to the 2D room-temperature FE α-In$_2$Se$_3$ [14-16]. Specifically, the application of a vertical electric field not only reverses the out-of-plane polarization but also induces a switch in the in-plane polarization direction. It is noteworthy that there exists a series of materials isostructural with AlAgS$_2$ that exhibit similar switchable ferroelectricity, such as the experimentally synthesized bulk LuCuS$_2$ and ScCuCS$_2$ [34,35] (see section 4 in SM). Moreover, type-I FQFE can exist not only in materials with polar point groups like AlAgS$_2$ but also in materials with non-polar point groups, such as LiAC$_2$ (A = Ag and Au) with *D$_{3h}$* point group (see section 5 in SM).

**Type-II FQFE in monolayer HgI$_2$.**

We now turn to investigate the monolayer HgI$_2$ that survives in the FQFE screening. Bulk HgI$_2$ is a van der Waals (vdWs) layered material that has also been synthesized experimentally [36,37]. The cleavage energy of monolayer $HgI_2$ is 0.21 $J/m^2$ (see Fig. S2), significantly lower than successfully exfoliated two-dimensional materials like graphene (0.46 $J/m^2$) [38] and $MoS_2$ (0.42 $J/m^2$) [39], suggesting that $HgI_2$ monolayers can be readily obtained through mechanical exfoliation. As depicted in Fig. 4a, the monolayer $HgI_2$ possesses a tetragonal lattice comprising one Hg atom and two I atoms. The $L_1$ phase of monolayer $HgI_2$ belongs to the *P-4m2* space group and has a non-polar point group of $D_{2d}$. Another phase $L_2$ can be constructed by *h* symmetry operation, where *h* can be inversion symmetry *I*, gliding mirror symmetry, four-fold rotation, two-fold screw, or mirror symmetry. Based on our symmetry strategy analysis, the $HgI_2$ monolayer is a candidate for FQFE candidate due to the fractional displacement of the Hg atoms are $\Delta d_{Hg,\alpha}$ = (1/2,1/2,0) (see Fig. 4a). However, due to the 2+ valence state of Hg ion, the system yields an integer polarization difference of $\Delta \mathbf{P} = \mathbf{Q}$ along the [110] direction, which demonstrate $HgI_2$ as a type-II FQFE.

We further demonstrated the FQFE nature of $HgI_2$ by DFT calculations. Firstly, we verify the thermal and dynamic stability by performing phonon and molecular dynamics simulations at 300 K for the low-symmetry phase. The absence of imaginary frequencies of phonons and the structure at the end of 300 K and 5 ps demonstrate the dynamically stable and the robustness of the structure under room-temperature, as shown in Fig. S3 in SM. Then, we explore the switching barrier from $L_1$ to $L_2$. We find that the barrier for $L_1$ to $L_2$ is 68 meV per Hg atom, which is much lower compared to popular traditional bulk FE materials like $PbTiO_3$ (200 meV/f.u.) [33], as shown in Fig. 4b. Then, we calculate the spontaneous polarization of the monolayer $HgI_2$ system using the Berry phase method [29-31]. As shown in Fig. 4(b), with an effective thickness of 6 Å, the spontaneous polarizations of $L_1$ and $L_2$ phases along the [110] direction are determined to be -42.0 and 42.0 $\mu C/cm^2$, respectively. Additionally, the polarization quantum of $HgI_2$ along the [110] direction is 84.0 $\mu C/cm^2$, according to the definition of $\mathbf{Q} = \frac{e}{\Omega}\mathbf{a}$ in modern polarization theory.

The polarization difference between two low-symmetry phase is $\Delta \mathbf{P} = \mathbf{Q}$, thus verifying that the monolayer $HgI_2$ is a type-II FQFE with an integer quantized polarization. These above results confirm that monolayer $HgI_2$ with non-polar point group possesses spontaneous and switchable type-II FQFE.

**Discussion.** Note that though we did not compute switching barriers for all the FQFE candidates in Table S1 and S2, there are still large possibility of experimental realization FQFE in those systems. This is due to that the barriers can decrease dramatically in presence of domain walls, strains, defects, etc [40-43]. For example, for the traditional ferroelectric $PbTiO_3$, the existence of 90° and 180° domain wall can lead to a minuscule barrier for reversal of polarization [40]. In addition, for the bulk $LiAuC_2$ with type-I FQFE, the lattice distortion can significantly reduce the barrier from 282 meV to 138 meV (see section 5 in SM). Furthermore, there are a significant number of materials predicted by DFT in the database that are expected to become candidates for FQFE. Using our proposed symmetry strategy, we identified 2759 materials predicted by DFT that have robust stability and exhibit FQFE properties. Among them, there are 411, 1062, and 1286 materials from Materials Project [23], C2DB [24,25], and GNoME database [44], respectively.

In summary, we develop a new method for identifying FQFE based on new symmetry principles. By integrating this method with high-throughput screening, we pinpoint 202 FQFE candidates, including 12 candidates within polar groups and 190 within non-polar groups. Through application of the FQFE theory and DFT calculations, we further validated the FQFE properties in $AlAgS_2$ and $HgI_2$ systems. Notably, we find that the $AlAgS_2$ exhibits an ultra-low switching barrier and its in-plane (FQFE) and out-of-plane (conventional FE) spontaneous polarizations measure 21.97 and 15.55 μC/cm², respectively. Additionally, monolayer $HgI_2$ exhibits large spontaneous polarization with 42 μC/cm². Our findings not only advance the understanding of FQFE but also provide various FQFE candidates for experimental verification.

**Acknowledgement**.

We acknowledge financial support from the National Key R&D Program of China (No. 2022YFA1402901), NSFC (No. 11825403, 11991061, 12188101, 12174060, and 12274082), the Guangdong Major Project of the Basic and Applied Basic Research (Future functional materials under extreme conditions—2021B0301030005), Shanghai Science and Technology Program (No. 23JC1400900), and Shanghai Pilot Program for Basic Research—FuDan University 21TQ1400100 (23TQ017). C.X. also acknowledges support from the Shanghai Science and Technology Committee (Grant No. 23ZR1406600) and the Xiaomi Young Talents Program.

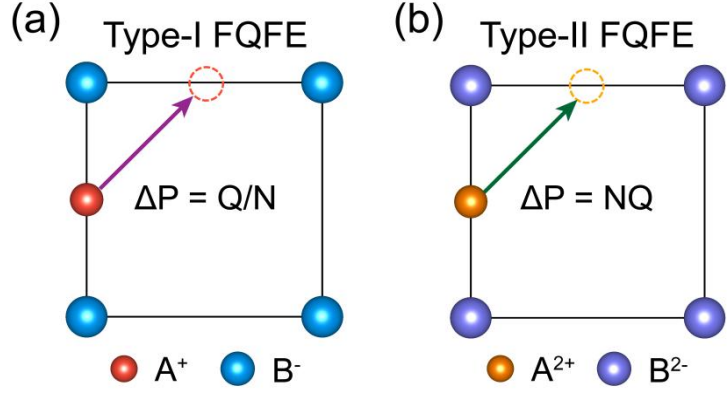

Fig1. Classification of FQFE. Schematics of (a) type-I FQFE and (b) type-II FQFE. The Type-I and Type-II FQFE systems exhibit ferroelectricity with fractional and integer quantized polarizations, respectively. The red and yellow balls represent movable and static ions, respectively. The arrows indicate the direction of movements, while dashed circles represent the destination of movements. The valence states of the ions are given at the bottom of the figure.

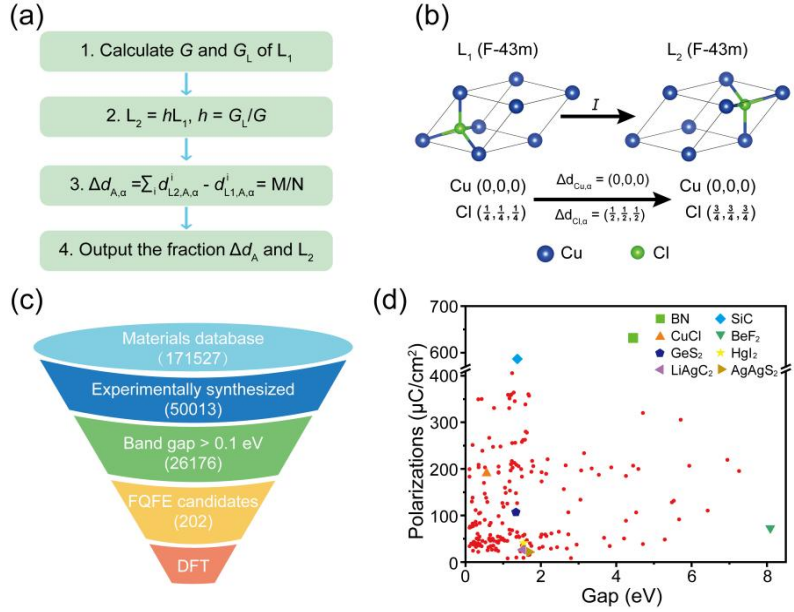

Figure 2. Symmetry strategy and high throughput screenings of the FQFE candidates. (a) The identification of FQFE candidates. $G$ is the space group of $L_1$ phase, while and $G_L$ is that of its lattice. $\Delta d_{A,\alpha}$ ($\alpha = x, y, z$) is the total displacement of the A atom(s) along the $\alpha$ direction in the phase transition from $L_1$-$L_2$, while $d^i_{L1,A,\alpha}$ and $d^i_{L2,A,\alpha}$ represent the α coordinates of the $i$-th A atom in the $L_1$ and $L_2$ phases. Note that $N$ and *M* choose their values from *N* = 2, 3, 4, 6, 8, *M* = 1, 2, ..., *N*-1. (b) FQFE in bulk CuCl with space group of the *F-43m*. (c) The high-throughput screening workflow of FQFE. (d) Spontaneous polarization versus band gap for FQFE candidates. The values of band gaps are extracted from the Materials Project [23] and C2DB [24,25].

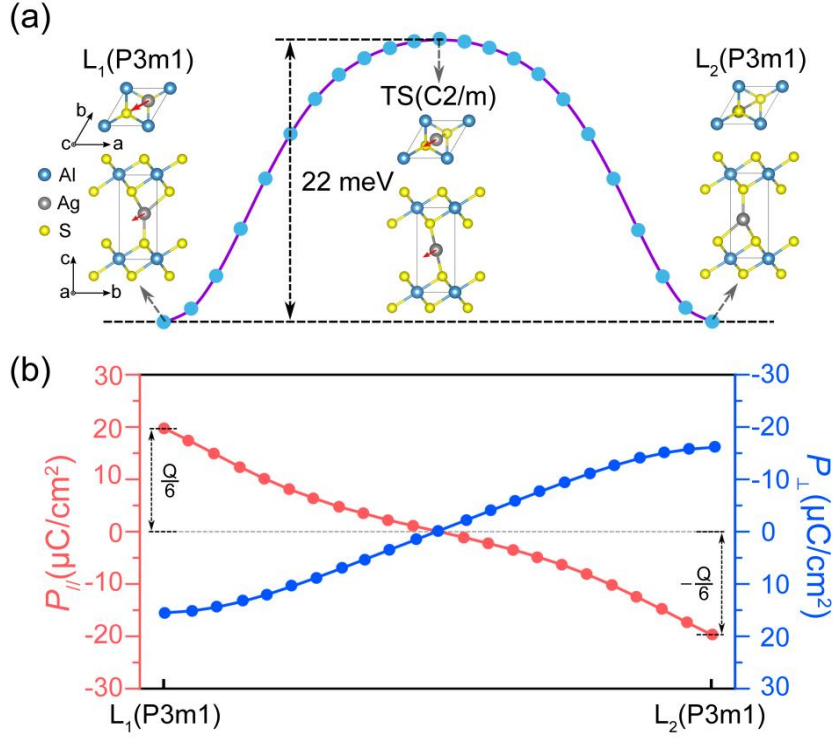

Figure 3. Type-I FQFE in bulk AlAgS$_2$. (a) Geometric structures and polarization switching barrier of AlAgS$_2$. The L$_{1/2}$ phase belongs to *P3m1* and the TS belongs to *C2/m*. The red arrow denotes the displacement of central Ag atoms, including an in-plane component of $\frac{1}{3}\mathbf{a} + \frac{1}{3}\mathbf{b}$ and a small out-of-plane displacement. (b) The evolution of in-plane and out-of-plane along the L$_1$-TS-L$_2$ pathway. The in-plane polarization $\mathbf{P}_\parallel$ of the L$_{1/2}$ phases are ±21.97 μC/cm$^2$ along the [110] direction. $\mathbf{Q}$ = 131.82 μC/cm$^2$ is the polarization quantum along the [110] direction. Thus, the $\mathbf{P}_\parallel$ = ±$\mathbf{Q}$/6 and the polarization difference between L$_1$ and L$_2$ phases is Δ$\mathbf{P}$ = $\mathbf{Q}$/3. The out-of-plane polarization $\mathbf{P}_\perp$ of the L$_{1/2}$ phases determined to be ±15.55 μC/cm$^2$.

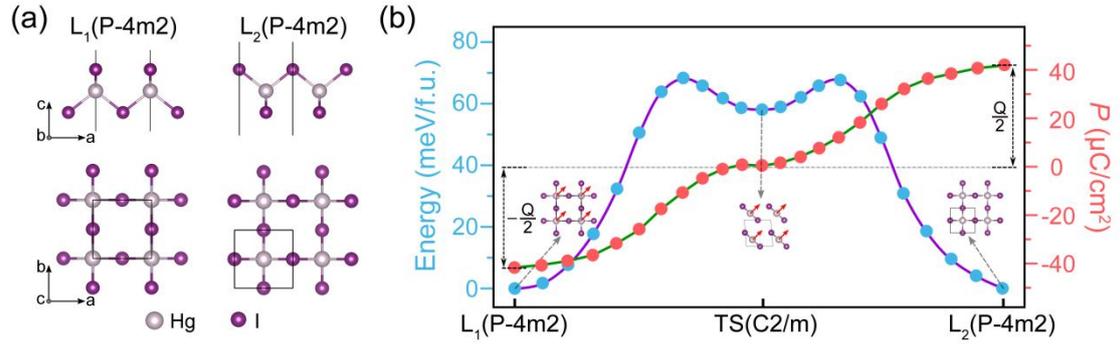

Figure 4. Type-II FQFE in monolayer HgI$_2$. (a) Geometric structure of HgI$_2$ for the L$_1$ and L$_2$ phases. (b) The energy barrier and the evolution of the polarization along the L$_1$-TS-L$_2$ pathways. The red arrows denote that the Hg atoms displace along the [110] direction. Here the polarizations of L$_{1/2}$ phases are **P**$_{1/2}$ = ±**Q**/2 = ±42.0 μC/cm$^2$. Thus, the polarization difference is Δ**P** = **Q**, which is an integer multiple of the polarization quantum. **Q** is the polarization quantum along the [110] direction.